\documentclass[aps,prl,superscriptaddress,twocolumn,nopacs,amsmath,showkeys,amssymb,letter]{revtex4}
\pdfoutput=1

\usepackage{color}
\usepackage{graphicx}
\usepackage{dcolumn}
\usepackage{bm}

\begin{document}

\title{Nearly-free-electron system of monolayer Na on the surface of single-crystal HfSe$_2$}

\author{T. Eknapakul}
\affiliation {School of Physics, Suranaree University of
Technology, Nakhon Ratchasima, 30000, Thailand}

\author{I. Fongkaew}
\affiliation {School of Physics, Suranaree University of
Technology, Nakhon Ratchasima, 30000, Thailand}

\author{S. Siriroj}
\affiliation {School of Physics, Suranaree University of
Technology, Nakhon Ratchasima, 30000, Thailand}

\author{R. Vidyasagar}
\affiliation {School of Physics, Suranaree University of Technology, Nakhon
Ratchasima, 30000, Thailand}

\author{J. D. Denlinger}
\affiliation {Advanced Light Source, Lawrence Berkeley National
Laboratory, Berkeley, CA 94720, USA}

\author{L. Bawden}
\affiliation {SUPA, School of Physics and Astronomy, University of St.
Andrews, St. Andrews, Fife KY16 9SS, United Kingdom}

\author{S.-K. Mo}
\affiliation {Advanced Light Source, Lawrence Berkeley National
Laboratory, Berkeley, CA 94720, USA}

\author{P. D. C. King}
\affiliation {SUPA, School of Physics and Astronomy, University of St.
Andrews, St. Andrews, Fife KY16 9SS, United Kingdom}

\author{H. Takagi}
\affiliation{Department of Physics, University of Tokyo, Hongo, Tokyo
113-0033, Japan} \affiliation{Magnetic Materials Laboratory, RIKEN Advanced
Science Institute, Wako, Saitama 351-0198, Japan}

\author{S. Limpijumnong}
\affiliation {School of Physics, Suranaree University of
Technology, Nakhon Ratchasima, 30000, Thailand} \affiliation
{NANOTEC-SUT Center of Excellence on Advanced Functional
Nanomaterials, Suranaree University of Technology, Nakhon
Ratchasima 30000, Thailand}

\author{W. Meevasana}
\altaffiliation {Corresponding e-mail: worawat@g.sut.ac.th}
\affiliation {School of Physics, Suranaree University of
Technology, Nakhon Ratchasima, 30000, Thailand} \affiliation
{NANOTEC-SUT Center of Excellence on Advanced Functional
Nanomaterials, Suranaree University of Technology, Nakhon
Ratchasima 30000, Thailand}

\date{\today}

\begin{abstract}
The electronic structure of a single Na monolayer on the surface of
single-crystal HfSe$_2$ is investigated using angle-resolved photoemission spectroscopy. We find that this
system exhibits an almost perfect ``nearly-free-electron'' behavior with an extracted effective
mass of $\sim\!1m_e$, in contrast to heavier masses found previously for alkali metal monolayers on other substrates.
Our density-functional-theory calculations indicate that this is due to the large lattice constant, causing both exchange
and correlation interactions to be suppressed, and to the weak
hybridization between the overlayer and the substrate. This is therefore an ideal model system for understanding the properties of two-dimensional materials.

\end{abstract}

\keywords{monolayer Na, weakly-interacting two-dimensional system, transition
metal dichalcogenide, HfSe$_2$,electronic structure, angle-resolved
photoemission spectroscopy.}
\maketitle

Understanding and controlling electrons in reduced dimensions, for example at the interfaces between disparate semiconductors, underpins modern electronic devices~\cite{convSi1,convSi2,convGaAs1,convGaAs2}. In recent years, this has found renewed prominence through the
study of electrons naturally confined in atomically-thin layers, such as in graphene or monolayer transition metal
dichalcogenides, opening prospects to achieve novel functionality such as ultrafast electronic~\cite{FETTgraphene}, spintronic, or valleytronic devices~\cite{spin1,spin2,spin3,spin4,valley1,valley2}. To progress towards these goals, it is critical to understand the behaviour of electrons in two-dimensional solids, and the influence of many-body interactions between them.

Angle-resolved photoemission spectroscopy (ARPES) is a powerful tool to achieve this. It directly
measures the electronic structure of materials, and can provide valuable information on carrier masses and the interactions between electrons in the system. This has been applied to numerous two-dimensional (or quasi-two-dimensional systems), including surface states of noble metals (e.g.
Cu~\cite{cu_felix,sos_cu}, Ag and Au~\cite{sos_au_ag,sos_au}),
semiconductors,~\cite{BiSe,MoSe2,MoS2,WSe2,Bi2Te3} and
metal oxides \cite{GaO, Phil_PtCoO2}, and alkali metals grown as two-dimensional layers on metallic
substrates~\cite{NaPt,NaNi, NaCu, NaCo,NaGraphite,NaBe}. Despite many of these systems being generally considered weakly interacting, there is
hardly any example of a system which displays true
nearly-free-electron behaviour manifested by a parabolic band dispersion with an effective carrier mass $m^*=1m_e$). For example, even for an alkali metal, in the cases of Na monolayers on Cu(111) \cite{NaCu} and Ni(100)
\cite{NaNi} surfaces, the effective mass was reported to be at least 30\% heavier than the bare
electron mass. This was attributed to a hybridization between the Na-derived electronic states and those of the underlying substrate. Moreover, exchange interactions can even lead to a lowering of the effective mass below unity~\cite{Li, tdeg1}, making achieving true free-electron-like behaviour very rare. In this work, we show how just such behaviour is manifested in a single Na layer stabilised on a semiconducting HfSe$_2$ substrate. We attribute this to particularly weak hybridization with the substrate due to a large out-of-plane lattice constant, and to a particularly weak exchange and correlation interaction.

\begin{figure*}
\includegraphics [width=7in, clip]{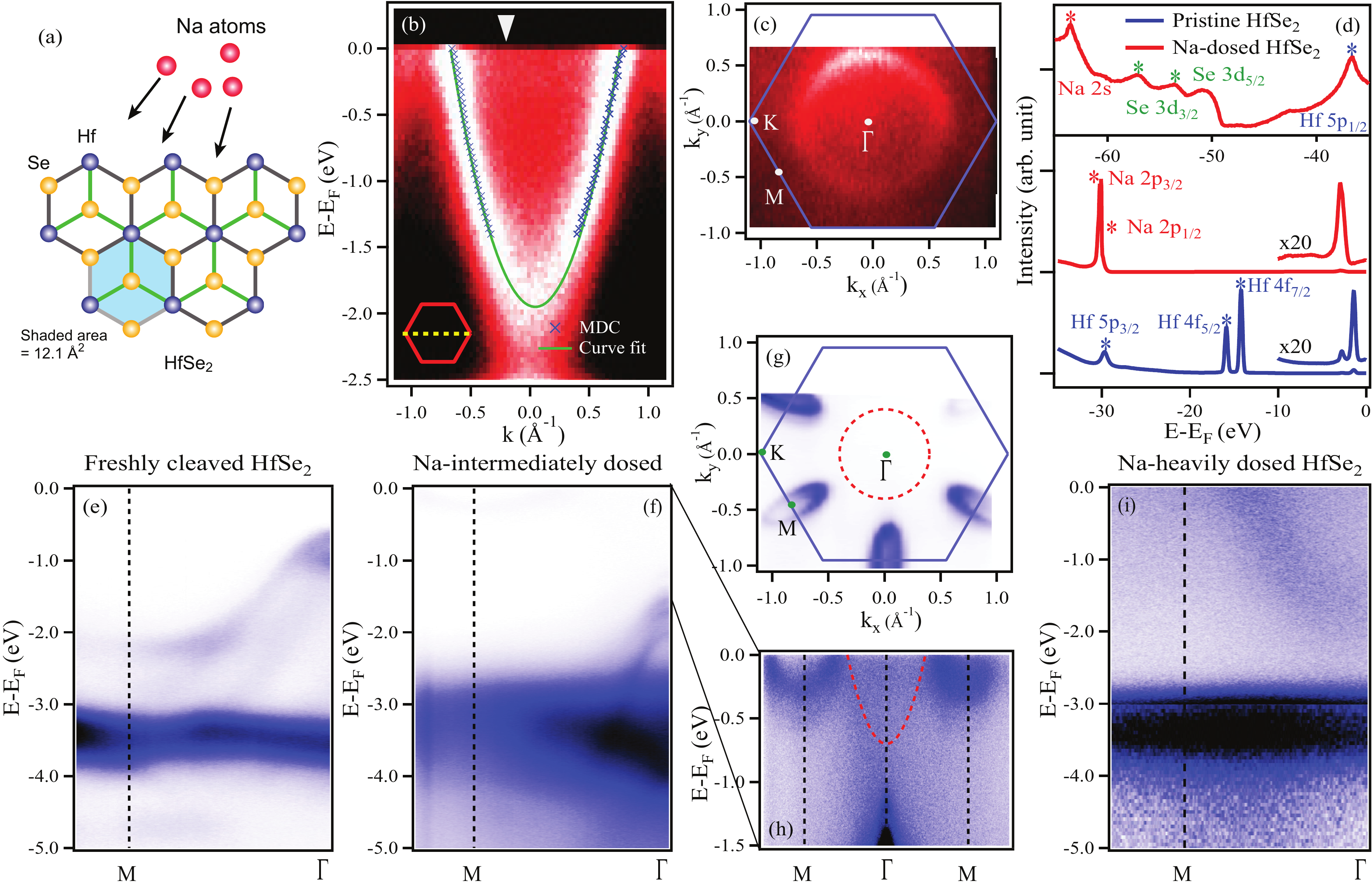}
\caption{\label{FIG1} (a) The atomic
structure of HfSe$_2$ surface (top view). (b) Parabolic band dispersion of
heavily-evaporated Na atoms on the surface of HfSe$_2$. (c)
Corresponding Fermi surface map of the band in (b).  (d) Angle-integrated
photoemission spectra of freshly-cleaved and Na-heavily-evaporated
HfSe$_2$, showing the core levels of Hf and Na (photon energy = 50 and 80 eV respectively). (e), (f) and (i) show
the valence bands of freshly-cleaved, intermediately-dosed (2 mins of dosing) and
heavily-dosed HfSe$_2$ (5 mins of dosing), respectively. (h) shows the zoom-in
data of the HfSe$_2$-band dispersion in panel (f) as indicated; dash line shows
a faint Na-band dispersion. (g) shows the Fermi surface map of intermediately-
dosed HfSe$_2$; note that this map is a different sample with similar dosing).}
\end{figure*}

1T-HfSe$_2$ single crystals, which we use as a substrate, were grown
using the flux method. This compound crystalises in the CdI$_2$ structure with a
hexagonal unit cell with in-plane and out-of-plane lattice constant of
a~=~3.74 $\AA$ and c~= 6.14 ~$\AA$ ~\cite{hflattice3,hflattice4},
respectively (see Fig. 1(a)). To obtain a clean surface, the crystal was
cleaved in ultra-high vacuum at a pressure better than 4$\times$10$^{-11}$
Torr. ARPES measurement were performed
immediately after cleaving, and following the deposition of sodium (Na) on the sample surface from a SAES
alkali metal source. The measurements were performed at beamlines 4.0.3 and 10.0.1 of the Advanced Light Source (USA)
using Scienta R4000 hemispherical electron analyzers. Photon energies were set to be in the range between 50 and 80 eV. The
sample temperature was maintained at between 40-80~K throughout the experiment.

Fig. 1(b)-(c) shows ARPES data measured after Na atoms were deposited
on the cleaved surface of single-crystal HfSe$_2$ for 5 mins. A dispersive
band with a parabolic shape is clearly observed. By fitting momentum
distribution curves (MDCs) of this parabolic band, we extracted the effective
mass to be $m^* = (1.00\pm0.04)m_e$ which is within error identical to the free
electron mass. The corresponding Fermi surface has a circular shape, again consistent with a free electron gas. From
this measured Fermi surface, we extract a surface carrier density from
the Luttinger area, $n_{2D}=k_F^2/2\pi=8.72\times10^{14}$~cm$^{-2}$.
We also note the possible presence of a second smaller band near E$_F$ (marked by an arrow in Fig. 1(b)) which is barely seen in the Na-heavily-dosed case. This is the strong confirmation of the well coverage of Na on HfSe$_2$ because our measured photon energies (50-80 eV) are very surface sensitive. We believe that this band is from the intercalation of Na atom into the vdW gap, however, from the weak intensity, the effect is likely not strong.

\begin{figure*}
\includegraphics [width=5.6in, clip]{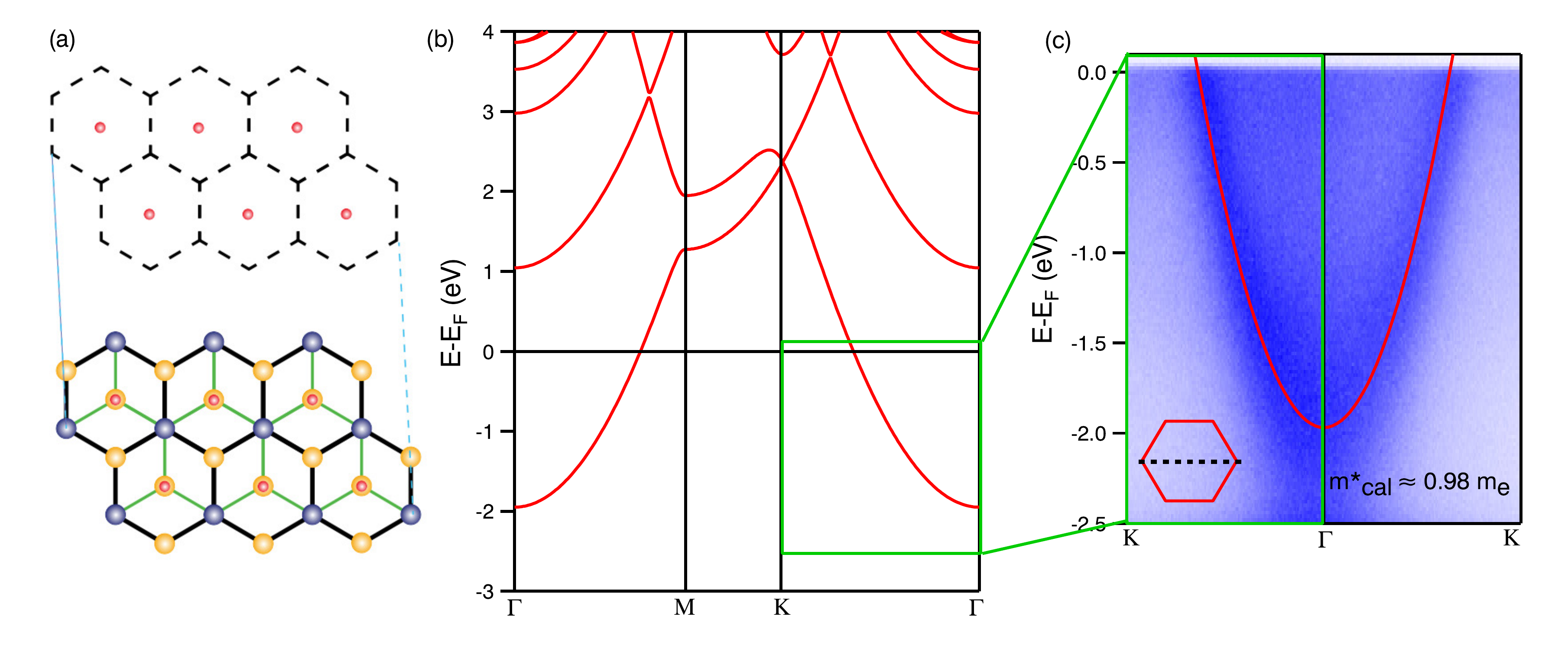}
\caption{\label{FIG2:sodium band} (a) Atomic structures of
monolayer Na on HfSe$_2$ (below) and bare monolayer Na (above)
which is used for calculating the band structure in (b). (c) shows
the zoom-in bands in the indicated box on top of the experimental
data from Fig. 1(a). Note that although there is some experimental
spectral weight of the smaller band in (c), it is much less clear
which may be due to that some small amount of Na atoms may get
intercalated to the inner HfSe$_2$ as in the case of MoS$_2$.}
\end{figure*}

Our observations of a dispersive band and a clearly-defined Fermi surface are indicative of a uniform and well-ordered metallic layer atop our semiconducting HfSe$_2$ substrate. To estimate the Na coverage, we perform additional measurements for shorter Na deposition times. As shown in Fig. 1(e)-(h), after depositing Na on the surface for 2 minutes, the HfSe$_2$ valence bands present in the freshly cleaved material (Fig.~1(e)) are still clearly visible, but shifted to higher binding energy concomitant with filling of the HfSe$_2$-derived conduction band states (charge transfer from the Na to the HfSe$_2$ populates the conduction band with carriers). The total electron density extracted from the Luttinger area of these conduction band pockets is estimated to be around
$3.8\times10^{14}$~cm$^{-2}$. Assuming a constant deposition rate, and assuming that each Na donates one electron, this would give a Na coverage of $9.4\times10^{14}$~atoms.cm$^{-2}$ for the 5 minute deposition. This is very close to the experimentally-extracted Luttinger count identified from the heavier dose above, implying that the results shown in Fig.~1(b,i) are from approximately 1.1~ML of Na on HfSe$_2$.

This is further confirmed by comparing the
angle-integrated core-level spectra before and after Na evaporation
as shown in Fig. 1(d). Before evaporation, Hf (4f$_{5/2}$ and
4f$_{7/2}$ ) peaks can be clearly observed at a binding energy of around 14-17
eV~\cite{hff,hflattice4}. After the heavy Na evaporation, Na (2p$_{3/2}$ and
2p$_{1/2}$) peaks at $\sim\!30$~eV binding energy become pronounced \cite{nap}
while the spectral weight of the Hf 4f peaks is almost completely suppressed. This confirms a uniform coverage of the Na overlayer. Additional Hf and Se-derived core levels are still observed at higher binding energy (e.g. between 35-60 eV, Fig. 1(d)). Due to the extreme surface sensitivity of photoemission performed at these energies, these results are entirely consistent with a single monolayer coverage of Na.

We now turn to our key observation that this Na monolayer hosts carriers with a effective mass so close to that of a free electron. In particular, this can be contrasted with similar systems such as a single
monolayer of Na on Cu(111) where the measured effective mass has been determined to be as heavy as 1.3$m_e$ \cite{NaCu}. This was suggested as a possible result of hybridization with the Cu substrate. We note that, in that case, the much smaller in-plane lattice
constant of Cu(111) (2.21 $\AA$ \cite{Cu111}) as compared to
bulk sodium ($a=3.77$ $\AA$) may cause a greater overlap of the wave functions of Na and Cu, implying such a hybridization can easily occur. Here, however, the in-plane lattice constant of the substrate (HfSe$_2$; $a=3.74$ $\AA$) is very
similar to that of Na lattice. Moreover, HfSe$_2$ is a layered material, dominated by weak van der Waals bonding between layers as in other transition
metal dichalcogenides~\cite{vdwn1,vdwn2,MoS2}. As such, hybridization between the Na and HfSe$_2$-derived states can be expected to be substantially weaker.

\begin{figure}
\includegraphics [width=3in, clip]{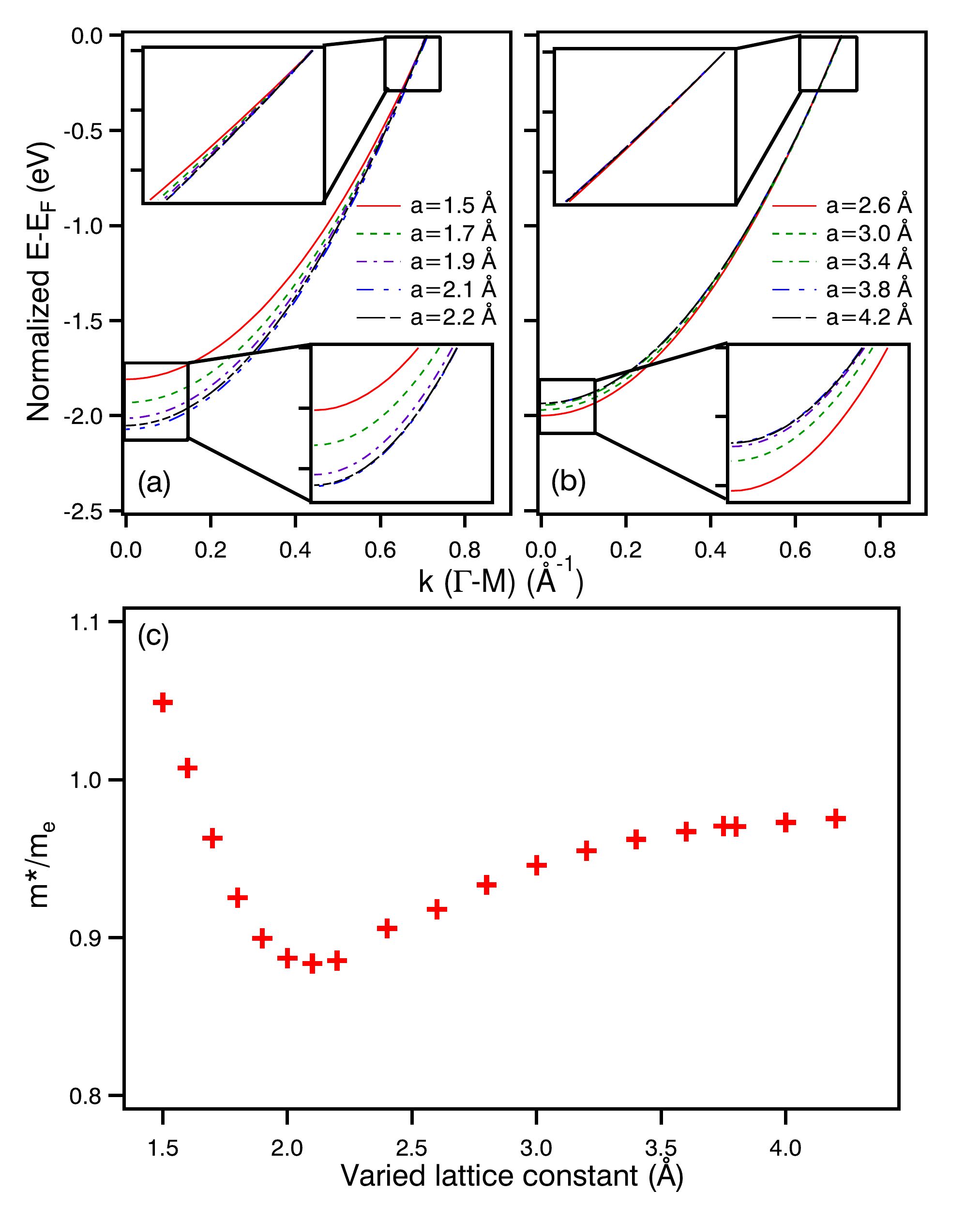}
\caption{\label{FIG4} (a) and (b) show calculated band dispersions of bare
monolayer Na with various lattice constants between 1.5 - 4.2 $\AA$. Insets show
the zoom-in bands at the Fermi level and the band bottoms. (c) shows
the effective masses extracted from the calculations in (a) and (b).}
\end{figure}

Based on this assumption of such a weak interaction,
we perform first-principles calculations of the band structure of an isolated Na monolayer (see Fig. 2). Previous low energy electron diffraction (LEED) studies show commensurate formations of Na atoms with the Na lattice reconstruction ranging from 3.125 to 5.412 $\AA$ on Cu(111) and Ru(0001) substrates~\cite{NaCuL,NaRuL1,NaRuL2}. This is a great support of our model for the (1$\times$1) structure due to the similar lattice constant between Na and HfSe$_2$.
The calculations were carried out
within the framework of density functional theory with projector augmented
wave potentials (PAW) \cite{cal1} as implemented in the VASP code. The PBE
approximation is used for the exchange correlation terms \cite{cal2,cal3}.
The electron wave functions were described using a plane wave basis set with
the energy cutoff of 520 eV. To calculate the 2D electronic band structures
of the Na monolayer, a periodic slab of monolayer Na[001] (P6$_3$/mmc
hexagonal structure) with 20~$\AA$  vacuum spacing between layers to prevent
interlayer interactions was used.  Positions of  Na atoms were relaxed until
the Hellmann-Feynman forces become less than 0.001eV/$\AA$ \cite{cal4} while
the in-plane cell vectors are kept at the theoretical relaxed bulk value a=
3.76 $\AA$ (expt. 3.77~$\AA$). For k-space integrations, we used the
Monkhorst-Pack scheme \cite{cal5} with 11 $\times$ 11 $\times$1 k-point
sampling.

The calculated bands are shown in Fig. 2(b). These show a clearly-dispersive band, which has an effective mass, m$^*_{cal}$ $\approx$~0.98$m_e$. This is in excellent agreement with our ARPES data, as shown in Fig. 2(c) where the calculations are overlaid on the data. This suggests that the larger lattice
constant of this substrate as compared to previously-investigated examples could be key in stabilising the free electron-like behavior in a monolayer of Na.
We note, also, that the high carrier densities can be expected to lead to very strong electronic screening, and as such the exchange interaction
(which can make the effective mass lighter) will remain small in comparison to the kinetic energy. This is in contrast to other two-dimensional electron gases (2DEGs) stabilised in much more poorly-screening materials, such as at the
interfaces/surfaces of oxides and low-doped dichalcogenides~\cite{Li,WSe2,tdeg1}.  We note that, as the lattice constant is reduced in our calculations (Fig. 3), and where screening can be expected to become less efficient, the effective mass of an isolated layer of Na can indeed become smaller than 1. At the smallest values, below 2.1 $\AA$, however, an increase in correlation energies cause the correlation term to dominate over exchange, leading to a steep increase in the effective mass.

Our findings show that, even for a simple model system of an alkali-metal single layer, there is possibility to engineer band structures via judicious choice of material substrates. In other words, a two-dimensional material cannot be considered in isolation. Its electronic structure can be influenced directly, via hybridization with the supporting medium, but also via many-body effects, e.g. by balancing and controlling the ratio of the exchange and correlation energies both to each other and to the kinetic energy. Understanding the fundamentals of this process will be key to designing desired properties in two-dimensional materials, e.g. negative electronic compressibility \cite{WSe2} and fast bandgap renormalization \cite{TRMoS2,PLbandgap}. Here we show that using a transition-metal dichalcogenide semiconductor as a support for a metallic Na single layer, we show that it is possible to minimize both of these effects, realising the unusual situation of an almost ideal nearly-free electron system with $m^*\sim$ 1$m_e$.

We acknowledge S. Chaiyachad, W. Jindata, C. Jaisuk,  Y. Kaekhamchan, P.
Chanprakhon for useful information. This work was supported by the SUT
Research and Development Fund, Suranaree University of Technology. The
Advanced Light Source is supported by the Office of Basic Energy Science of
the US DOE under contract No. DE-AC02-05CH11231.

\end{document}